\documentclass[preprint,nofootinbib]{revtex4}%
\usepackage{amssymb}
\usepackage{amsfonts}
\usepackage{amsmath}
\usepackage{graphicx}
\usepackage[usenames]{color}%
\providecommand{\U}[1]{\protect\rule{.1in}{.1in}}
\definecolor{blue}{rgb}{0,0,1}

\definecolor{red}{rgb}{1,0,0}

\begin{document}
\title{All the solutions of the form $M_{2}\times_{W}\Sigma_{d-2}$ for Lovelock
gravity in vacuum in the Chern-Simons case}
\author{Julio Oliva}
\affiliation{Instituto de Ciencias F\'{\i}sicas y Matem\'{a}ticas, Universidad Austral de
Chile, Valdivia, Chile\footnote{julio.oliva@docentes.uach.cl}}
\affiliation{Universidad de Buenos Aires, FCEN-UBA, Ciudad Universitaria, Pabell\'{o}n I,
1428, Buenos Aires, Argentina.}

\begin{abstract}
In this note we classify a certain family of solutions of Lovelock gravity in
the Chern-Simons (CS) case, in arbitrary (odd) dimension, $d\geq5$. The
spacetime is characterized by admitting a metric that is a warped product of a
two-dimensional spacetime $M_{2}$ and an (a priori) arbitrary Euclidean
manifold $\Sigma_{d-2}$ of dimension $d-2$. We show that the solutions are
naturally classified in terms of the equations that restrict $\Sigma_{d-2}$.
According to the strength of such constraints we found the following branches
in which $\Sigma_{d-2}$ has to fulfill: a Lovelock equation with a single
vacuum (Euclidean Lovelock Chern-Simons in dimension $d-2$), a single scalar
equation that is the trace of an Euclidean Lovelock CS equation in dimension
$d-2$, or finally a degenerate case in which $\Sigma_{d-2}$ is not restricted
at all. We show that all the cases have some degeneracy in the sense that the
metric functions are not completely fixed by the field equations. This result
extends the static five-dimensional case previously discussed in Phys.Rev. D76
(2007) 064038, and it shows that in the CS case, the inclusion of higher
powers in the curvature does not introduce new branches of solutions in
Lovelock gravity. Finally we comment on how the inclusion of a non-vanishing
torsion may modify this analysis.

\end{abstract}
\maketitle

\section{Introduction}

Gravity in higher dimensions has proved to be an interesting arena to test how
generic are the notions gained in four dimensional gravitational physics. Even
in higher dimensional General Relativity (GR), properties as uniqueness and
stability of solutions in vacuum may depart completely from their
four-dimensional counterpart (for a recent summary of the state of the art see
\cite{Horowitzbook}). Maintaining the second order character of the field
equations in higher dimensions, it is possible to consider a more general
setup than the one defined by Einstein's gravity, since as proved by Lovelock
in \cite{Lovelock} the most general parity-even Lagrangian in arbitrary
dimension $d$, that gives second order field equations for the metric is given
by an arbitrary linear combination of the dimensional continuations of all the
lower dimensional Euler densities. This gives rise to the so-called Lovelock
gravity, the simplest case after GR being the Einstein-Gauss-Bonnet (EGB)
gravity. In this theory, in addition to the Einstein-Hilbert and cosmological
terms, one includes a term which is quadratic in the curvature and gives
non-trivial field equations in dimensions grater than four. This quadratic
combination is very precise, in such a way that the possible higher derivative
terms cancel each other and one gets second order field equations. Since the
field equations come from a diffeomorphism invariant action, their divergence
vanishes identically.

To find exact and analytic solutions of these theories is a non-trivial
problem when one departs from spherical symmetry\footnote{Departing from the
family of metrics (\ref{ans}), looking for exact rotating solutions is also a
more difficult than in GR, since for example considering the Kerr-Schild
ansatz that naturally gives rise to the Myers-Perry solution with cosmological
constant in GR \cite{rotgr}, one finds that in order to have a non-trivial
solution in EGB, the coupling constants must be fixed as in the Chern-Simons
case \cite{rot} and even more the solution turns out to be non-circular
\cite{rot2}, making the analysis of the causal structure more cumbersome (for
some perturbative and numerical solutions see also \cite{otherrot}).}. For
example, a problem that is solved in a very simple manner in GR, corresponds
to finding the most general solution of the form%
\begin{equation}
ds_{d}^{2}=-f^{2}\left(  t,r\right)  dt^{2}+\frac{dr^{2}}{g^{2}\left(
t,r\right)  }+r^{2}d\Sigma_{d-2}^{2}\ . \label{ans}%
\end{equation}
where $\Sigma_{d-2}$ is an arbitrary Euclidean manifold of dimension $d-2$.
Einstein equations plus a cosmological constant in vacuum%
\begin{equation}
G_{\mu\nu}+\Lambda g_{\mu\nu}=0\ ,
\end{equation}
imply that the metric functions do not depend on $t$, and are given by%
\begin{equation}
f^{2}=g^{2}=-\frac{2\Lambda r^{2}}{\left(  d-1\right)  \left(  d-2\right)
}-\frac{\mu}{r^{d-2}}+\gamma\ ,
\end{equation}
where $\mu$ is an arbitrary integration constant and $\Sigma_{d-2}$ must be an
Einstein manifold fulfilling the equation%
\begin{equation}
\tilde{R}_{ij}=\left(  d-3\right)  \gamma\tilde{g}_{ij}\ .
\end{equation}
Here $\tilde{R}_{ij}$ is the Ricci tensor of $\Sigma_{d-2}$ and $\tilde
{g}_{ij}$ its metric \cite{GRbase}.

Solving exactly the same problem in Lovelock gravity is more complicated. For
example, in the EGB theory for the static case, the work \cite{DOT3} solves
this problem in arbitrary dimension finding a rich set of causal structures.
For arbitrary values of the coupling constants of the theory, the analysis
done in \cite{DOT3} reduces to the done previously reported in
\cite{DottiGleiser}, where it was proved that if one assumes $\Sigma_{d-2}$ to
be Einstein, then one can show that it must also obey a quadratic restriction
on the Weyl tensor which includes a new parameter $\theta$. That parameter
appears in the lapse function and even more, it modifies the asymptotic
behavior of the metric (see also \cite{Hidekidg}).

For arbitrary $\Sigma$, beyond the EGB case not much is known. The static
solution in the spherically symmetric case was found in \cite{JTWheeler}. When
$\Sigma_{d-2}$ is a constant curvature manifold, a Birkhoff's theorem was
proved in \cite{Zegers} (see also \cite{DeserFranklin}). Reference
\cite{Zegers} also shows that Birkhoff's theorem is not valid when the
coupling constants are fixed in a precise way and some degeneracies may appear
since in such cases, some of the metric functions are not determined by the
field equations (for some particular cases, this was previously observed in
reference \cite{Charmousis}). Lovelock theory, being a gravity theory with
higher powers in the curvature, could have more than one maximally symmetric
solution, and the mentioned degeneracies appear precisely at the regions in
the space of couplings in which some of these vacua coincide\footnote{See also
reference \cite{IR} for some solutions of the EGB theory in the case in which
there is no maximally symmetric solution at all.} (for some static black hole
solutions, with constant curvature horizons in this case see \cite{BHSs}).

It would be interesting therefore to classify all the solutions of the form
(\ref{ans}) in higher curvature Lovelock theories. In this work we focus on
the odd-dimensional case, when the highest possible power of the curvature is
present in the Lagrangian and all the vacua coincide. This theory is known as
Lovelock-Chern-Simons (LCS) theory (for a recent review see \cite{reviewZ}).

The action for a general Lovelock theory can be written as%
\begin{equation}
I=\kappa\int%
{\displaystyle\sum\limits_{p=0}^{\left[  \frac{d-1}{2}\right]  }}
\alpha_{p}\varepsilon_{a_{1}...a_{2p}a_{2p+1}...a_{d}}\overset{p-times}%
{\overbrace{R^{a_{1}a_{2}}...R^{a_{2p-1}a_{2p}}}}e^{a_{2p+1}}...e^{a_{d}}\ ,
\label{action}%
\end{equation}
where $\kappa$ and $\alpha_{p}$ are arbitrary (dimensionfull) coupling
constants, $\varepsilon_{a_{1}...a_{d}}$ is the Lorentz invariant Levi-Civita
tensor, $R^{ab}:=d\omega^{ab}+\omega^{ac}\omega_{c}^{\ b}\ $is the curvature
two-form written in terms Lorentz connection one-form $\omega^{ab}$, and
$e^{a}$ is the vielbein. $\left[  x\right]  $ stands for the integer part of
$x$. Wedge exterior product between differential forms is understood. Finally,
latin indices $\left\{  a_{i},b_{i}\right\}  $ run from $0$ to $d-1$.

The term with $p=0$ in (\ref{action}), corresponds to a volume term that gives
the contribution of the cosmological constant, for $p=1$ one gets the
Einstein-Hilbert term, while for $p=2$ the Lagrangian reduces to the
Gauss-Bonnet term. As mentioned before, here we will focus on the case
$d=2n+1$ and the coefficients $\alpha_{p}$ are given by%
\begin{equation}
\alpha_{p}:=\frac{1}{2n-2p+1}\binom{n}{p}\frac{1}{l^{2\left(  n-p\right)  }%
}\ , \label{coeff}%
\end{equation}
where $l^{2}$ is the squared curvature radius of the unique ($AdS$) maximally
symmetric solution. For simplicity we will focus on the case $l^{2}>0$,
nevertheless the de Sitter case is trivially obtained by analytically
continuing $l\rightarrow il$, while the flat limit (up to some subtleties that
will be mentioned\ when necessary) can be obtained by taking $l\rightarrow
\infty$ .

When torsion vanishes, the field equations coming from (\ref{action}) with the
couplings given by (\ref{coeff}) can be written as%
\begin{equation}
E_{a}:=\varepsilon_{aa_{1}...a_{2n}}\overset{n-times}{\overbrace{\bar
{R}^{a_{1}a_{2}}...\bar{R}^{a_{2n-1}a_{2n}}}}=0\ , \label{feqforms}%
\end{equation}
where we have defined the concircular curvature two-form as $\bar{R}%
^{ab}:=R^{ab}+\frac{1}{l^{2}}e^{a}e^{b}$. In terms of tensors, if we use the
generalized Kronecker delta of strength one denoted by $\delta_{\beta
_{1}...\beta_{p}}^{\alpha_{1}...\alpha_{p}}$, by defining the concircular
curvature tensor $\bar{R}_{\ \ \gamma\delta}^{\alpha\beta}=R_{\ \ \gamma
\delta}^{\alpha\beta}+\frac{1}{l^{2}}\delta_{\gamma\delta}^{\alpha\beta}$, the
field equations (\ref{feqforms}) read%
\begin{equation}
E_{\ \beta}^{\alpha}:=\delta_{\beta\beta_{1}...\beta_{2n}}^{\alpha\alpha
_{1}...\alpha_{2n}}\overset{n-times}{\overbrace{\bar{R}_{\ \ \ \alpha
_{1}\alpha_{2}}^{\beta_{1}\beta_{2}}...\bar{R}_{\ \ \ \alpha_{2n-1}\alpha
_{2n}}^{\beta_{2n-1}\beta_{2n}}}}=0\ . \label{feqtensors}%
\end{equation}

In the next section we will prove that all the solutions of the form
(\ref{ans}), for the field equation (\ref{feqforms}) (or equivalently
(\ref{feqtensors})) fall into one of the following three different classes:

\bigskip

\textbf{Case I:} The manifold $\Sigma_{d-2}$ is arbitrary and the metric reads%
\begin{equation}
ds^{2}=-\left(  \frac{r^{2}}{l^{2}}-\mu\right)  dt^{2}+\frac{dr^{2}}%
{\frac{r^{2}}{l^{2}}-\mu}+r^{2}d\Sigma_{d-2}^{2}\ ,
\end{equation}
where $\mu$ is an integration constant.

\bigskip

\textbf{Case II:} For $\xi\neq0$, if the manifold $\Sigma_{d-2}$ satisfies the
following (scalar) restriction%
\begin{equation}
\varepsilon_{i_{1}...i_{2n-2}}\overset{\left(  n-1\right)  -times}%
{\overbrace{\left(  \tilde{R}^{i_{1}i_{2}}-\xi e^{i_{1}i_{2}}\right)
...\left(  \tilde{R}^{i_{2n-3}i_{2n-2}}-\xi e^{i_{2n-3}i_{2n-2}}\right)  }%
}=0\ , \label{case2bm}%
\end{equation}
where $\tilde{R}^{ij}$ is the curvature two-form intrinsically defined on
$\Sigma_{d-2}$ and the indices $\left\{  i,j\right\}  $ run on $\Sigma_{d-2}$,
then the metric reads%
\begin{equation}
ds^{2}=-\left(  c_{1}\left(  t\right)  r+c_{2}\left(  t\right)  \sqrt
{\frac{r^{2}}{l^{2}}+\xi}\right)  ^{2}dt^{2}+\frac{dr^{2}}{\frac{r^{2}}{l^{2}%
}+\xi}+r^{2}d\Sigma_{d-2}^{2}\ ,
\end{equation}
with $c_{1}\left(  t\right)  $ and $c_{2}\left(  t\right)  $ arbitrary
integration functions. In the flat limit ($l\rightarrow\infty$) the metric
reduces to%
\[
ds^{2}=-\left(  c_{1}\left(  t\right)  r+c_{2}\left(  t\right)  \right)
^{2}dt^{2}+\frac{dr^{2}}{\xi}+r^{2}d\Sigma_{d-2}^{2}\ ,
\]

In the case $\xi=0$ (which does not exist in the limit $l\rightarrow\infty$)
the restriction on $\Sigma_{d-2}$ is obtained by setting $\xi=0$ in
(\ref{case2bm}) and the metric reads%
\begin{equation}
ds^{2}=-\left(  c_{1}\left(  t\right)  r+\frac{c_{2}\left(  t\right)  }%
{r}\right)  ^{2}dt^{2}+\frac{l^{2}dr^{2}}{r^{2}}+r^{2}d\Sigma_{d-2}^{2}\ ,
\end{equation}
where again $c_{1}\left(  t\right)  $ and $c_{2}\left(  t\right)  $ are
arbitrary integration functions. Note that in all of these cases, by
redefining the time coordinate, one can gauge away one of the two integration
functions, but not both simultaneously.

\bigskip

\textbf{Case III:} The manifold $\Sigma_{d-2}$ satisfies the following tensor
restriction%
\begin{equation}
\varepsilon_{ji_{1}...i_{2n-2}}\overset{\left(  n-1\right)  -times}%
{\overbrace{\left(  \tilde{R}^{i_{1}i_{2}}-\xi e^{i_{1}i_{2}}\right)
...\left(  \tilde{R}^{i_{2n-3}i_{2n-2}}-\xi e^{i_{2n-3}i_{2n-2}}\right)  }%
}\ =0\ ,
\end{equation}
and the metric reads%
\begin{equation}
ds^{2}=-f^{2}\left(  t,r\right)  dt^{2}+\frac{dr^{2}}{\frac{r^{2}}{l^{2}}+\xi
}+r^{2}d\Sigma_{d-2}^{2}\ ,
\end{equation}
with $f\left(  t,r\right)  $ an arbitrary function.

\bigskip

This result extends the static five dimensional case previously analyzed in
\cite{DOT1}. In Case I, we see that the manifold $\Sigma_{d-2}$ is arbitrary,
i.e. it is not fixed by the field equations. In Case II, the manifold
$\Sigma_{d-2}$ is fixed by a single scalar equation which, even after using
diffeomorphism invariance, in general it is not enough to determine a metric
on it. Finally in Case III, we see that the lapse function $f^{2}\left(
t,r\right)  $ is left arbitrary by the field equations. Therefore we conclude
that, in the previously mentioned sense, all the cases have some degeneracy.

\section{Proof of the classification}

To develop the proof of the classification it is useful to have the components
of the curvature two-form with respect to some basis for the metric
(\ref{ans}). If we define the components of the vielbein as%
\begin{equation}
e^{0}=fdt\text{, }e^{1}=\frac{dr}{g}\text{ and }e^{i}=r\tilde{e}^{i}\ ,
\end{equation}
where $\tilde{e}^{i}$ is the vielbein intrinsically defined on $\Sigma_{d-2}$,
then the nontrivial components of the concircular curvature two-form $\bar
{R}^{ab}$ read%
\begin{equation}
\bar{R}^{01}=Ae^{0}e^{1}\text{, }\bar{R}^{0i}=Be^{0}e^{i}+Ce^{1}e^{i}\text{,
}\bar{R}^{1i}=Fe^{1}e^{i}+He^{0}e^{i}\text{ and }\bar{R}^{ij}=\tilde{R}%
^{ij}+Je^{i}e^{j}\ ,
\end{equation}
where $\tilde{R}^{ij}$ is the curvature two-form intrinsically defined on
$\Sigma_{d-2}$ and $A$, $B$, $C$, $F$, $H$ and $J$ are functions of $t$ and
$r$ defined by%
\begin{align}
A  &  =A(t,r):=-\frac{g}{f}\left[  \left(  \frac{\dot{g}}{g^{2}f}\right)
^{\cdot}+\left(  gf^{\prime}\right)  ^{\prime}\right]  +\frac{1}{l^{2}%
}\ ,\label{compr1}\\
B  &  =B(t,r):=-g^{2}\frac{f^{\prime}}{rf}+\frac{1}{l^{2}}\ ,\ C:=C(t,r)=\frac
{\dot{g}}{fr}\\
F  &  =F(t,r):=-\frac{\left(  g^{2}\right)  ^{\prime}}{2r}+\frac{1}{l^{2}%
}\ ,\ H:=H\left(  t,r\right)  =-\frac{\dot{g}}{rf}\\
J  &  =J(t,r):=-\frac{g^{2}}{r^{2}}+\frac{1}{l^{2}}\ . \label{compr4}%
\end{align}
Primes denote derivation with respect to $r$ while dots derivation with
respect to $t$.

There are three kinds of equations depending on whether the free index in
(\ref{feqforms}) goes along the time direction, radial direction or along the
manifold $\Sigma_{d-2}$, which respectively reduce to%
\begin{align*}
E_{0}  &  :=2n\varepsilon_{01i_{1}...i_{2n-1}}\bar{R}^{1i_{1}}\overset
{n-1-times}{\overbrace{\bar{R}^{i_{2}i_{3}}...\bar{R}^{i_{2n-2}i_{2n-1}}}%
}=0\ ,\\
E_{1}  &  :=2n\varepsilon_{10i_{1}...i_{2n-1}}\bar{R}^{0i_{1}}\overset
{n-1-times}{\overbrace{\bar{R}^{i_{2}i_{3}}...\bar{R}^{i_{2n-2}i_{2n-1}}}%
}=0\ ,\\
E_{j}  &  :=2n\varepsilon_{j01i_{1}...i_{2n-2}}\bar{R}^{01}\overset
{n-1-times}{\overbrace{\bar{R}^{i_{1}i_{2}}...\bar{R}^{i_{2n-3}i_{2n-2}}}%
}+2n\left(  2n-2\right)  \varepsilon_{j0i_{1}1i_{2}i_{3}...i_{2n-2}}\bar
{R}^{0i_{1}}\bar{R}^{1i_{2}}\overset{n-2-times}{\overbrace{\bar{R}^{i_{3}%
i_{4}}...\bar{R}^{i_{2n-3}i_{2n-2}}}}\ =0\ .
\end{align*}
After introducing explicitly in these equations the components of the
concircular curvature two-form (\ref{compr1})-(\ref{compr4}), we get the
following three equations%
\begin{align*}
\mathcal{G}_{0}  &  :=\left(  Fe^{1}e^{i_{1}}+He^{0}e^{i_{1}}\right)
\varepsilon_{01i_{1}...i_{2n-1}}\overset{n-1-times}{\overbrace{\left(
\tilde{R}^{i_{2}i_{3}}+Jr^{2}\tilde{e}^{i_{2}}\tilde{e}^{i_{3}}\right)
...\left(  \tilde{R}^{i_{2n-2}i_{2n-1}}+Jr^{2}e^{i_{2n-2}}e^{i_{2n-1}}\right)
}}=0\ ,\\
\mathcal{G}_{1}  &  :=\left(  Be^{0}e^{i_{1}}+Ce^{1}e^{i_{1}}\right)
\varepsilon_{01i_{1}...i_{2n-1}}\underset{n-1-times}{\underbrace{\left(
\tilde{R}^{i_{2}i_{3}}+Jr^{2}\tilde{e}^{i_{2}}\tilde{e}^{i_{3}}\right)
...\left(  \tilde{R}^{i_{2n-2}i_{2n-1}}+Jr^{2}e^{i_{2n-2}}e^{i_{2n-1}}\right)
}}=0\ ,
\end{align*}
and%
\begin{gather*}
\mathcal{G}_{j}:=\left.  A\varepsilon_{ji_{1}...i_{2n-2}}\overset
{n-1-times}{\overbrace{\left(  \tilde{R}^{i_{1}i_{2}}+Jr^{2}\tilde{e}^{i_{1}%
}\tilde{e}^{i_{2}}\right)  ...\left(  \tilde{R}^{i_{2n-3}i_{2n-2}}%
+Jr^{2}e^{i_{2n-3}}e^{i_{2n-2}}\right)  }}\right. \\
\left.  +2\left(  n-1\right)  \left(  BF-CH\right)  r^{4}\varepsilon
_{ji_{1}...i_{2n-2}}\tilde{e}^{i_{1}}\tilde{e}^{i_{2}}\overset{n-2-times}%
{\overbrace{\left(  \tilde{R}^{i_{3}i_{4}}+Jr^{2}\tilde{e}^{i_{3}}\tilde
{e}^{i_{4}}\right)  ...\left(  \tilde{R}^{i_{2n-3}i_{2n-2}}+Jr^{2}e^{i_{2n-3}%
}e^{i_{2n-2}}\right)  }}\right.  =0\ ,
\end{gather*}
where we have defined $\varepsilon_{i_{1}...i_{2n-1}}:=\varepsilon
_{01i_{1}...i_{2n-1}}$.

Considering the combinations $e^{0}\mathcal{G}_{0}+e^{1}\mathcal{G}_{1}=0$ and
$e^{1}\mathcal{G}_{0}+e^{0}\mathcal{G}_{1}=0$ one respectively gets%
\begin{align}
\left(  F-B\right)  \varepsilon_{01i_{1}...i_{2n-1}}\overset{n-1-times}%
{\overbrace{\left(  \tilde{R}^{i_{1}i_{2}}+Jr^{2}\tilde{e}^{i_{1}}\tilde
{e}^{i_{2}}\right)  ...\left(  \tilde{R}^{i_{2n-3}i_{2n-2}}+Jr^{2}e^{i_{2n-3}%
}e^{i_{2n-2}}\right)  }\tilde{e}^{i_{2n-1}}}  &  =0\ ,\label{g0menosg1}\\
\left(  H-C\right)  \varepsilon_{01i_{1}...i_{2n-1}}\overset{n-1-times}%
{\overbrace{\left(  \tilde{R}^{i_{1}i_{2}}+Jr^{2}\tilde{e}^{i_{1}}\tilde
{e}^{i_{2}}\right)  ...\left(  \tilde{R}^{i_{2n-3}i_{2n-2}}+Jr^{2}e^{i_{2n-3}%
}e^{i_{2n-2}}\right)  }\tilde{e}^{i_{2n-1}}}  &  =0\ . \label{g0menosg12}%
\end{align}
This immediately splits the analysis in two cases defined by the (would be)
constraint on $\Sigma_{d-2}$%
\begin{equation}
\varepsilon_{i_{1}...i_{2n-1}}\overset{n-1-times}{\overbrace{\left(  \tilde
{R}^{i_{1}i_{2}}+Jr^{2}\tilde{e}^{i_{1}}\tilde{e}^{i_{2}}\right)  ...\left(
\tilde{R}^{i_{2n-3}i_{2n-2}}+Jr^{2}\tilde{e}^{i_{2n-3}}\tilde{e}^{i_{2n-2}%
}\right)  }\tilde{e}^{i_{2n-1}}}=0\ . \label{basescalarthetar}%
\end{equation}

If (\ref{basescalarthetar}) doesn't hold, then we need to impose $F=B$ and
$H=C$, the former implies that $g\left(  t,r\right)  =g\left(  r\right)  $,
while the later implies $f\left(  t,r\right)  =S(t)g\left(  r\right)  $. The
function $S\left(  t\right)  $ can be set to $1$ without lost of generality by
means of a redefinition of the time coordinate. Therefore in this branch (i.e.
provided (\ref{basescalarthetar}) doesn't hold), we have that (\ref{g0menosg1}%
) and (\ref{g0menosg12}) imply $f(t,r)=g\left(  t,r\right)  =f\left(
r\right)  =g\left(  r\right)  $. If (\ref{basescalarthetar}) holds, then
$\mathcal{G}_{0}=0=\mathcal{G}_{1}$ without imposing any restriction on the
function $f$ and $g$ at the moment. Note that the quantities with tilde on top
depend only on the coordinates in $\Sigma_{d-2}$, while the combination
$Jr^{2}$, could depend on both $t$ and $r$. At the moment this is not relevant
since equations (\ref{g0menosg1}) and (\ref{g0menosg12}) are factorized in any
case, but later we will see that the consistency of equation
(\ref{basescalarthetar}) strongly constraints the metric functions.

If we consider now equation $\mathcal{G}_{0}=0$, in the case in which
(\ref{basescalarthetar}) doesn't hold and therefore $f^{2}\left(  r\right)
=g^{2}\left(  r\right)  $ then we can see that $H$ identically vanishes, while
the vanishing of the function $F$ implies that $g^{2}=\frac{r^{2}}{l^{2}}-\mu
$, where $\mu$ is an integration constant. As mentioned, in this branch we
also have $f^{2}=\frac{r^{2}}{l^{2}}-\mu$ (since $f^{2}=g^{2}$), and therefore
one can see by direct evaluation that $A$ identically vanishes also. Therefore
$H=F=A=0$ and then equation $\mathcal{G}_{i}=0$ is also trivially satisfied
without imposing any restriction on $\Sigma_{d-2}$. \textbf{This concludes the
proof of Case I outlined in the introduction.}

On the other hand if (\ref{basescalarthetar}) holds (as mentioned before)
$\mathcal{G}_{0}$ and $\mathcal{G}_{1}$ vanish identically and then at the
moment, the functions $f\left(  t,r\right)  $ and $g\left(  t,r\right)  $ are
not restricted. Before continuing to equation $\mathcal{G}_{i}=0$, let us go
back to the problem of the consistency of equation (\ref{basescalarthetar}).
Considering the derivative of this equation with respect to the $t$ and $r$,
we respectively obtain%
\begin{align}
\left(  n-1\right)  \frac{\partial\left(  Jr^{2}\right)  }{\partial
r}\varepsilon_{i_{1}...i_{2n-1}}\overset{n-2-times}{\overbrace{\left(
\tilde{R}^{i_{1}i_{2}}+Jr^{2}\tilde{e}^{i_{1}}\tilde{e}^{i_{2}}\right)
...\left(  \tilde{R}^{i_{2n-5}i_{2n-4}}+Jr^{2}\tilde{e}^{i_{2n-5}}\tilde
{e}^{i_{2n-4}}\right)  }}\tilde{e}^{i_{2n-3}}\tilde{e}^{i_{2n-2}}\tilde
{e}^{i_{2n-1}}  &  =0\ ,\label{prs}\\
\left(  n-1\right)  \frac{\partial\left(  Jr^{2}\right)  }{\partial
t}\varepsilon_{i_{1}...i_{2n-1}}\underset{n-2-times}{\underbrace{\left(
\tilde{R}^{i_{1}i_{2}}+Jr^{2}\tilde{e}^{i_{1}}\tilde{e}^{i_{2}}\right)
...\left(  \tilde{R}^{i_{2n-5}i_{2n-4}}+Jr^{2}\tilde{e}^{i_{2n-5}}\tilde
{e}^{i_{2n-4}}\right)  }}\tilde{e}^{i_{2n-3}}\tilde{e}^{i_{2n-2}}\tilde
{e}^{i_{2n-1}}  &  =0\ , \label{pts}%
\end{align}
therefore $\left(  Jr^{2}\right)  ^{\prime}=\left(  Jr^{2}\right)  ^{\cdot}=0$
and consequently $Jr^{2}=-\xi$ with $\xi$ a constant, or the second term in
both (\ref{prs}) and (\ref{pts}) vanishes, implying a new scalar restriction
on $\Sigma_{d-2}$ that contains terms of order $n-2$ in the curvature and
might also depend on $t,r$, therefore its compatibility must be analyzed as
well. The first case ($Jr^{2}=-\xi$), implies $g^{2}\left(  t,r\right)
=g^{2}\left(  r\right)  =\frac{r^{2}}{l^{2}}+\xi$ where $\xi$ is an
integration constant. Note also that when $n=2$, we are forced to set $\left(
Jr^{2}\right)  ^{\prime}=\left(  Jr^{2}\right)  ^{\cdot}=0$ otherwise the
volume element on $\Sigma_{d-2}$ would vanish. On the other hand (for $n>2$),
if we assume $\left(  Jr^{2}\right)  ^{\prime}$ and $\left(  Jr^{2}\right)
^{\cdot}$ to be nonvanishing we can divide these factors obtaining the new
mentioned scalar restriction on $\Sigma_{d-2}$, which reads%
\begin{equation}
\varepsilon_{i_{1}...i_{2n-1}}\overset{n-2-times}{\overbrace{\left(  \tilde
{R}^{i_{1}i_{2}}+Jr^{2}\tilde{e}^{i_{1}}\tilde{e}^{i_{2}}\right)  ...\left(
\tilde{R}^{i_{2n-5}i_{2n-4}}+Jr^{2}\tilde{e}^{i_{2n-5}}\tilde{e}^{i_{2n-4}%
}\right)  }}\tilde{e}^{i_{2n-3}}\tilde{e}^{i_{2n-2}}\tilde{e}^{i_{2n-1}}=0\ .
\end{equation}
Again, we must consider the consistency of this equation by taking its
derivative with respect to the parameters $r$ and $t$. This respectively gives%
\begin{align}
\left(  n-2\right)  \frac{\partial\left(  Jr^{2}\right)  }{\partial
r}\varepsilon_{i_{1}...i_{2n-1}}\overset{n-3-times}{\overbrace{\left(
\tilde{R}^{i_{1}i_{2}}+Jr^{2}\tilde{e}^{i_{1}}\tilde{e}^{i_{2}}\right)
...\left(  \tilde{R}^{i_{2n-7}i_{2n-6}}+Jr^{2}\tilde{e}^{i_{2n-7}}\tilde
{e}^{i_{2n-6}}\right)  }}\tilde{e}^{i_{2n-5}}...\tilde{e}^{i_{2n-1}}  &
=0\ ,\\
\left(  n-2\right)  \frac{\partial\left(  Jr^{2}\right)  }{\partial
t}\varepsilon_{i_{1}...i_{2n-1}}\underset{n-3-times}{\underbrace{\left(
\tilde{R}^{i_{1}i_{2}}+Jr^{2}\tilde{e}^{i_{1}}\tilde{e}^{i_{2}}\right)
...\left(  \tilde{R}^{i_{2n-7}i_{2n-6}}+Jr^{2}\tilde{e}^{i_{2n-7}}\tilde
{e}^{i_{2n-6}}\right)  }}\tilde{e}^{i_{2n-5}}...\tilde{e}^{i_{2n-1}}  &  =0\ .
\end{align}
If $n=3$ we are forced again to set $\left(  Jr^{2}\right)  ^{\prime}=\left(
Jr^{2}\right)  ^{\cdot}=0$ (otherwise the volume element of $\Sigma_{d-2}$
should vanish) which fixes $g^{2}=\frac{r^{2}}{l^{2}}+\xi$, while for $n>3$ we
can consider $\left(  Jr^{2}\right)  ^{\prime}$ and $\left(  Jr^{2}\right)
^{\cdot}$ to be nonvanishing and divide by these expressions, therefore
obtaining another scalar restriction on $\Sigma_{d-2}$, which this time,
includes powers of the curvature of order $n-3$. Repeating this procedure
$n-1$ times one eventually gets%
\begin{align}
\frac{\partial\left(  Jr^{2}\right)  }{\partial r}\varepsilon_{i_{1}%
...i_{2n-1}}\tilde{e}^{i_{1}}...\tilde{e}^{i_{2n-1}}  &  =0\ ,\\
\frac{\partial\left(  Jr^{2}\right)  }{\partial t}\varepsilon_{i_{1}%
...i_{2n-1}}\tilde{e}^{i_{1}}...\tilde{e}^{i_{2n-1}}  &  =0\ ,
\end{align}
and if the expressions $\left(  Jr^{2}\right)  ^{\prime}$ and $\left(
Jr^{2}\right)  ^{\cdot}$ are nonvanishing then we would have that the volume
form of $\Sigma_{d-2}$ must vanish, arriving to a contradiction. \textbf{We
have proved then that in the case }$F\neq B$\textbf{ and }$H\neq C$,\textbf{
the consistency of equation (\ref{basescalarthetar}) implies that }$\left(
Jr^{2}\right)  ^{\prime}=\left(  Jr^{2}\right)  ^{\cdot}=0$\textbf{ which in
turn implies that }$g^{2}=\frac{r^{2}}{l^{2}}+\xi$\textbf{ with }$\xi$\textbf{
an integration constant, and consequently (\ref{basescalarthetar}) reads}%
\begin{equation}
\varepsilon_{i_{1}...i_{2n-1}}\overset{n-1-times}{\overbrace{\left(  \tilde
{R}^{i_{1}i_{2}}-\xi\tilde{e}^{i_{1}}\tilde{e}^{i_{2}}\right)  ...\left(
\tilde{R}^{i_{2n-3}i_{2n-2}}-\xi\tilde{e}^{i_{2n-3}}\tilde{e}^{i_{2n-2}%
}\right)  }\tilde{e}^{i_{2n-1}}}=0\ , \label{scalarnmenos1sigma}%
\end{equation}
which now depends only on the coordinates of $\Sigma_{d-2}$.

The remaining structure comes from the analysis of equation $\mathcal{G}%
_{i}=0$. Note that $g^{2}=\frac{r^{2}}{l^{2}}+\xi$ further implies $H=0=F$,
therefore $\mathcal{G}_{j}$ reduces to%
\begin{equation}
A\varepsilon_{ji_{1}...i_{2n-2}}\overset{n-1-times}{\overbrace{\left(
\tilde{R}^{i_{1}i_{2}}-\xi\tilde{e}^{i_{1}}\tilde{e}^{i_{2}}\right)
...\left(  \tilde{R}^{i_{2n-3}i_{2n-2}}-\xi e^{i_{2n-3}}e^{i_{2n-2}}\right)
}}\ =0\ . \label{gi}%
\end{equation}
If $\xi\neq0$, the equation $A=0$ allows to integrate $f\left(  t,r\right)  $,
which in this case reads%
\begin{equation}
f^{2}=\left(  c_{1}\left(  t\right)  r+c_{2}\left(  t\right)  \sqrt
{\frac{r^{2}}{l^{2}}+\xi}\right)  ^{2}\ \text{,}%
\end{equation}
while for $\xi=0$ it integrates as%
\begin{equation}
f^{2}=\left(  c_{1}\left(  t\right)  r+\frac{c_{2}\left(  t\right)  }%
{r}\right)  ^{2}\ .
\end{equation}
The latter case is not defined in the flat limit ($l\rightarrow\infty$) while
in such a limit, when $g^{2}=\xi$, the equation $A=0$ gives the following
expression for $f$:%
\[
f^{2}\left(  t,r\right)  =\left(  c_{1}\left(  t\right)  r+c_{2}\left(
t\right)  \right)  ^{2}\ .
\]

In all of these expressions $c_{1}\left(  t\right)  $ and $c_{2}\left(
t\right)  $ are arbitrary integration functions and note that one of them can
be gauged away by a redefinition of the time coordinate.\bigskip

Summarizing, in this branch we have that if $\xi\neq0$, the metric reads%
\begin{equation}
ds^{2}=-\left(  c_{1}\left(  t\right)  r+c_{2}\left(  t\right)  \sqrt
{\frac{r^{2}}{l^{2}}+\xi}\right)  ^{2}dt^{2}+\frac{dr^{2}}{\frac{r^{2}}{l^{2}%
}+\xi}+r^{2}d\Sigma_{d-2}^{2}\ ,
\end{equation}
which in the limit $l\rightarrow\infty$ takes the form%
\[
ds^{2}=-\left(  c_{1}\left(  t\right)  r+c_{2}\left(  t\right)  \right)
^{2}dt^{2}+\frac{dr^{2}}{\frac{r^{2}}{l^{2}}+\xi}+r^{2}d\Sigma_{d-2}^{2}\ ,
\]
while for $\xi=0$ we have
\[
ds^{2}=-\left(  c_{1}\left(  t\right)  r+\frac{c_{2}\left(  t\right)  }%
{r}\right)  ^{2}dt^{2}+\frac{l^{2}dr^{2}}{r^{2}}+r^{2}d\Sigma_{d-2}^{2}\ ,
\]
where $c_{1}\left(  t\right)  $ and $c_{2}\left(  t\right)  $ are arbitrary
integration functions and $\Sigma_{d-2}$ fulfills in both cases, the same
scalar equation (\ref{scalarnmenos1sigma}). Note that here the constant $\xi$
appears in the restriction on $\Sigma_{d-2}$ and can be scaled to $\pm1$ when
it is non-vanishing. \textbf{This ends the proof of Case 2 outlined in the
introduction.}

If $A\neq0$, equation (\ref{gi}) implies a tensor restriction on $\Sigma
_{d-2}$, which naturally, is stronger than its trace given by
(\ref{scalarnmenos1sigma}). When this tensor restriction holds, the metric
reads%
\[
ds^{2}=-f^{2}\left(  t,r\right)  dt^{2}+\frac{dr^{2}}{\frac{r^{2}}{l^{2}}+\xi
}+r^{2}d\Sigma_{d-2}^{2}\ ,
\]
with $\Sigma_{d-2}$ constrained by%
\[
\varepsilon_{ji_{1}...i_{2n-2}}\overset{n-1-times}{\overbrace{\left(
\tilde{R}^{i_{1}i_{2}}-\xi\tilde{e}^{i_{1}}\tilde{e}^{i_{2}}\right)
...\left(  \tilde{R}^{i_{2n-3}i_{2n-2}}-\xi e^{i_{2n-3}}e^{i_{2n-2}}\right)
}}\ =0\ ,
\]
and the function $f\left(  t,r\right)  $ is arbitrary. \textbf{This concludes
the proof of Case 3 outlined in the introduction. }This tensor restriction
corresponds to an Euclidean Lovelock CS equation in dimension $d-2=2n-1$.

\bigskip

This concludes the proof of the classification.

\section{Discussion}

\textbf{On the causal structures}

For the Case 2 and Case 3, the $(t,r)$-part of the metrics obtained depend on
arbitrary functions of the time coordinate, therefore the causal structure of
this spacetimes is not fixed. Note that this dependence cannot be gauged away
completely by a diffeomorphism. Nevertheless, a few comments on the causal
structures are in order in all of the three cases when the integration
functions are chosen to be constants, i.e. $c_{1}\left(  t\right)  =c_{1}$ and
$c_{2}\left(  t\right)  =c_{2}$.

In Case I, the solution describes a black hole. This solution reduces to the
one found in \cite{CSBTZ}. In such case also, its thermodynamics and causal
structure coincide with that of the three-dimensional
Ba\~{n}ados-Teitelboim-Zanelli (BTZ) black hole \cite{BTZ} where, for generic
values of $\mu$, the causal structure singularity at $r=0$ of the
three-dimensional case is now replaced by a curvature singularity as can be
seen by evaluating, for example, the Ricci scalar.

In Case II, the metric
\begin{equation}
ds^{2}=-\left(  c_{1}r+c_{2}\sqrt{\frac{r^{2}}{l^{2}}+\xi}\right)  ^{2}%
dt^{2}+\frac{dr^{2}}{\frac{r^{2}}{l^{2}}+\xi}+r^{2}d\Sigma_{d-2}^{2}\ ,
\end{equation}
might describe the traversable wormhole found in \cite{DOTWorm}, which is
asymptotically AdS at both asymptotic regions. This is the case when $\xi=-1$
and $|\frac{c_{2}}{lc_{1}}|<1$, which can be seen directly by performing the
change of coordinates $r=l\cosh\rho$ and allowing the coordinate $\rho$ to go
from $-\infty$ to $+\infty$. In this case, the metric reduces to%
\begin{equation}
ds^{2}=l^{2}\left[  -\cosh^{2}\left(  \rho-\rho_{0}\right)  dt^{2}+d\rho
^{2}+\cosh^{2}\rho d\Sigma_{d-2}^{2}\right]  \ ,
\end{equation}
where $\rho_{0}=-\tanh^{-1}\left(  \frac{c_{2}}{lc_{1}}\right)  $ and we have
properly rescaled the time coordinate. The conditions under which the
propagation of a scalar field on this background is stable, was studied in
\cite{COT}, and some holographic properties of strings attached to the
boundaries have been explored in \cite{StringsWorm}. For $\xi=0$ the metric
reduces to%
\begin{equation}
ds^{2}=-\left(  c_{1}r+\frac{c_{2}}{r}\right)  ^{2}dt^{2}+\frac{l^{2}dr^{2}%
}{r^{2}}+r^{2}d\Sigma_{d-2}^{2}\ .
\end{equation}
When $c_{1}\neq0$ this spacetime is asymptotically locally AdS, while if
$c_{1}=0$, the $(t,r)$-part of the metric reduces to a Lifshitz geometry
(geometry with an anisotropic scaling symmetry), with a dynamic exponent
equals to $z=-1$.

Since in Case III the lapse function is arbitrary, the causal structure is
also undefined even in the static case.

\bigskip

\textbf{Does torsion help removing the degeneracy?}

The field equations coming from the variation with respect to the spin
connection in Lovelock theory, do not necessarily imply that torsion should
vanish (for some explicit solutions see e.g. \cite{torsols}). For example in
five dimensions, in first order formalism for the Lovelock CS case, the field
equations coming from the variation with respect to the vielbein and the spin
connection are respectively given by%
\begin{align}
\varepsilon_{abcde}\left(  R^{bc}+\frac{1}{l^{2}}e^{b}e^{c}\right)  \left(
R^{de}+\frac{1}{l^{2}}e^{d}e^{e}\right)   &  =0\label{feqtor1}\\
\varepsilon_{abcde}\left(  R^{cd}+\frac{1}{l^{2}}e^{c}e^{d}\right)  T^{e}  &
=0\ , \label{feqtor2}%
\end{align}
where we have introduced the torsion two-form $T^{e}:=De^{e}:=\frac{1}%
{2}e_{\ \alpha}^{e}T_{\ \mu\nu}^{\alpha}dx^{\mu}\wedge dx^{\nu}$. Therefore
choosing the Levi-Civita connection is ad-hoc. Then it is natural to wonder
whether the equations coming from the torsion may help removing the
degeneracy. Posing the question in a different manner one could ask : is there
a non-degenerate branch of solutions of (\ref{feqtor1})-(\ref{feqtor2}) in
which the vielbein and the spin connection are compatible with the local
isometries of $\Sigma_{d-2}$?. It is clear that there are particular cases in
which the torsion may not be vanishing and anyway the system is degenerated
since, if for example we choose the (non-Riemannian) curvature to be constant
$R^{ab}=-\frac{1}{l^{2}}e^{a}e^{b}$, then the torsion is left completely
arbitrary by the field equations. Note also that, since this theory has an
extra symmetry that mixes the spin connection and the vielbein (see
\cite{reviewZ}), the arbitrariness in the torsion can be transformed into an
arbitrariness of the line element constructed out from the corresponding
vielbein. A thorough analysis with the inclusion of torsion will be presented
elsewhere \cite{Thor}.

\bigskip

\textbf{Further comments}

As studied for the static quadratic case in \cite{DOT3}, when one considers
Lovelock theories that do not belong to the subclass of Lovelock CS, but
nevertheless the couplings are related in such a way that there is a unique
vacuum, there are also sectors in which some of the metric functions are
arbitrary. Therefore this phenomenon seems to be more related to the fact of
having degenerate maximally symmetric solution than with the appearance of an
extra symmetry. In such non-Lovelock CS theories, as well as in the Lovelock
CS\ ones, this degeneracy allows to have interesting causal structures as
solutions (see e.g. \cite{Javier}). Nevertheless in the former cases, there
are more restrictions on $\Sigma_{d-2}$, which on one hand can be thought of
as helping to remove the degeneracy, while in the other hand could be not
compatible beyond the constant curvature case. A simple set of geometries
beyond constant curvature manifolds (or their products) are product of the
homogenous three dimensional Thurston geometries, which have been recently
found to provide simple examples of transverse sections of hairy black holes
for some Lovelock theories in even dimensions \cite{hairylovelock}. In the
context of compactifications of Lovelock CS theories, involving metrics that
are products of constant curvature spaces, the degenerate behavior is also
present as it was proved in reference \cite{MH} back in the early $90$'s. The
inclusion of matter fields seems to help removing the mentioned degeneracies
(see for example references \cite{nondeg}).

If one departs from the underlying $(A)dS$ symmetry group, static spherically
symmetric solutions of gravitational CS theories with matter fields, have also
been recently considered in \cite{QS}. In this reference, the authors
considered a Chern-Simons theory evaluated on a Lie algebra that is obtained
by performing what the authors called an $S$-expansion procedure
\cite{Sexpansion} from the $AdS$ algebra and a particular semigroup $S$, which
provides an approach to obtain GR in odd-dimensions from a CS theory. It would
be interesting to study further the properties of these theories and to
integrate them in the general ansatz (\ref{ans}) classifying the possible
non-degenerate sectors.

\bigskip

\section{Acknowledgments}

We thank Andr\'{e}s Anabal\'{o}n, Fabrizio Canfora, Francisco Correa, Gustavo
Dotti, Sourya Ray and Steven Willison for many useful conversations. We thank
also the support of Becas Chile Postdoctorales, CONICYT, 2012 and FONDECYT
grant 11090281.


\bigskip

\begin{thebibliography}{99}                                                                                               %


\bibitem {Horowitzbook}Gary T. Horowitz (Editor), "Black Holes in Higher
Dimensions", Cambridge University Press; 1 edition (May 28, 2012), ISBN 1107013453.

\bibitem {Lovelock}D. Lovelock,
J.\ Math.\ Phys.\ \textbf{12}, 498 (1971)

\bibitem {GRbase}D.~Birmingham,
Class.\ Quant.\ Grav.\ \textbf{16}, 1197 (1999) [hep-th/9808032].
G.~Gibbons and S.~A.~Hartnoll,
Phys.\ Rev.\ D \textbf{66}, 064024 (2002) [hep-th/0206202].
G.~W.~Gibbons, S.~A.~Hartnoll and C.~N.~Pope,
Phys.\ Rev.\ D \textbf{67}, 084024 (2003) [hep-th/0208031].


\bibitem {rotgr}S.~W.~Hawking, C.~J.~Hunter and M.~Taylor,
Phys.\ Rev.\ D \textbf{59}, 064005 (1999) [hep-th/9811056].
G.~W.~Gibbons, H.~Lu, D.~N.~Page and C.~N.~Pope,
J.\ Geom.\ Phys.\ \textbf{53}, 49 (2005) [hep-th/0404008].


\bibitem {rot}A.~Anabalon, N.~Deruelle, Y.~Morisawa, J.~Oliva, M.~Sasaki,
D.~Tempo and R.~Troncoso,
Class.\ Quant.\ Grav.\ \textbf{26}, 065002 (2009) [arXiv:0812.3194 [hep-th]].


\bibitem {rot2}A.~Anabalon, N.~Deruelle, D.~Tempo and R.~Troncoso,
Int.\ J.\ Mod.\ Phys.\ D \textbf{20}, 639 (2011) [arXiv:1009.3030 [gr-qc]].


\bibitem {otherrot}H.~-C.~Kim and R.~-G.~Cai,
Phys.\ Rev.\ D \textbf{77}, 024045 (2008) [arXiv:0711.0885 [hep-th]].
Y.~Brihaye and E.~Radu,
Phys.\ Lett.\ B \textbf{661}, 167 (2008) [arXiv:0801.1021 [hep-th]].
B.~Kleihaus, J.~Kunz, E.~Radu and B.~Subagyo,
Phys.\ Lett.\ B \textbf{713}, 110 (2012) [arXiv:1205.1656 [gr-qc]].
Y.~Brihaye, B.~Kleihaus, J.~Kunz and E.~Radu,
JHEP \textbf{1011}, 098 (2010) [arXiv:1010.0860 [hep-th]].


\bibitem {DOT3}G.~Dotti, J.~Oliva, R.~Troncoso,
Phys.\ Rev.\ \textbf{D82}, 024002 (2010). [arXiv:1004.5287 [hep-th]].

\bibitem {DottiGleiser}G.~Dotti, R.~J.~Gleiser,
Phys.\ Lett.\ \textbf{B627}, 174-179 (2005). [hep-th/0508118].

\bibitem {Hidekidg}H.~Maeda,
Phys.\ Rev.\ D \textbf{81}, 124007 (2010) [arXiv:1004.0917 [gr-qc]].


\bibitem {JTWheeler}J.~T.~Wheeler,
Nucl.\ Phys.\ B \textbf{273}, 732 (1986).


\bibitem {Zegers}R.~Zegers,
J.\ Math.\ Phys.\ \textbf{46}, 072502 (2005) [gr-qc/0505016].


\bibitem {DeserFranklin}S.~Deser and J.~Franklin,
Class.\ Quant.\ Grav.\ \textbf{22}, L103 (2005) [gr-qc/0506014].


\bibitem {Charmousis}C.~Charmousis and J.~-F.~Dufaux,
Class.\ Quant.\ Grav.\ \textbf{19}, 4671 (2002) [hep-th/0202107].


\bibitem {IR}F.~Izaurieta and E.~Rodriguez,
arXiv:1207.1496 [hep-th].


\bibitem {BHSs}J.~Crisostomo, R.~Troncoso and J.~Zanelli,
Phys.\ Rev.\ D \textbf{62}, 084013 (2000) [hep-th/0003271].
R.~Aros, R.~Troncoso and J.~Zanelli,
Phys.\ Rev.\ D \textbf{63}, 084015 (2001) [hep-th/0011097].


\bibitem {reviewZ}J.~Zanelli,
Class.\ Quant.\ Grav.\ \textbf{29}, 133001 (2012) [arXiv:1208.3353 [hep-th]].


\bibitem {DOT1}G.~Dotti, J.~Oliva, R.~Troncoso,
Phys.\ Rev.\ \textbf{D76}, 064038 (2007). [arXiv:0706.1830 [hep-th]].

\bibitem {CSBTZ}M. Banados, C. Teitelboim and J. Zanelli,
Phys.\ Rev.\ D \textbf{49}, 975 (1994).

\bibitem {BTZ}M. Banados, C. Teitelboim and J. Zanelli,
Phys.\ Rev.\ Lett.\ \textbf{69}, 1849 (1992).

\bibitem {DOTWorm}G.~Dotti, J.~Oliva and R.~Troncoso,
Phys.\ Rev.\ D \textbf{75}, 024002 (2007) [hep-th/0607062].


\bibitem {COT}D.~H.~Correa, J.~Oliva and R.~Troncoso,
JHEP \textbf{0808}, 081 (2008) [arXiv:0805.1513 [hep-th]].


\bibitem {StringsWorm}M.~Ali, F.~Ruiz, C.~Saint-Victor and
J.~F.~Vazquez-Poritz,
Phys.\ Rev.\ D \textbf{80}, 046002 (2009) [arXiv:0905.4766 [hep-th]].
M.~Ali, F.~Ruiz, C.~Saint-Victor and J.~F.~Vazquez-Poritz,
arXiv:1005.5541 [hep-th].
R.~E.~Arias, M.~Botta Cantcheff and G.~A.~Silva,
Phys.\ Rev.\ D \textbf{83}, 066015 (2011) [arXiv:1012.4478 [hep-th]].


\bibitem {torsols}F.~Canfora, A.~Giacomini and S.~Willison,
Phys.\ Rev.\ D \textbf{76}, 044021 (2007) [arXiv:0706.2891 [gr-qc]].
F.~Canfora, A.~Giacomini and R.~Troncoso,
Phys.\ Rev.\ D \textbf{77}, 024002 (2008) [arXiv:0707.1056 [hep-th]].
F.~Canfora and A.~Giacomini,
Phys.\ Rev.\ D \textbf{82}, 024022 (2010) [arXiv:1005.0091 [gr-qc]].


\bibitem {Thor}F. Canfora, A. Giacomini and J. Oliva. Work in progress.

\bibitem {Javier}J.~Matulich and R.~Troncoso,
JHEP \textbf{1110}, 118 (2011) [arXiv:1107.5568 [hep-th]].


\bibitem {hairylovelock}A.~Anabalon, F.~Canfora, A.~Giacomini and J.~Oliva,
Phys.\ Rev.\ D \textbf{84}, 084015 (2011) [arXiv:1108.1476 [hep-th]].


\bibitem {MH}F.~Mueller-Hoissen,
Nucl.\ Phys.\ B \textbf{346}, 235 (1990).


\bibitem {nondeg}O.~Miskovic, R.~Troncoso and J.~Zanelli,
Phys.\ Lett.\ B \textbf{615}, 277 (2005) [hep-th/0504055].
;M.~H.~Dehghani and R.~B.~Mann,
JHEP \textbf{1007}, 019 (2010) [arXiv:1004.4397 [hep-th]].


\bibitem {QS}C. A. C. Quinzacara, P. Salgado, Phys. Rev. D \textbf{85}, 124026 (2012).

\bibitem {Sexpansion}F.~Izaurieta, E.~Rodriguez and P.~Salgado,
J.\ Math.\ Phys.\ \textbf{47}, 123512 (2006) [hep-th/0606215].
J.\ Math.\ Phys.\ \textbf{50}, 073511 (2009) [arXiv:0903.4712 [hep-th]].

\end{thebibliography}
\end{document}